\documentclass[review,runningheads]{llncs}
\usepackage[dvipsnames]{xcolor}
\usepackage{amsmath}
\usepackage[breaklinks=true]{hyperref}
\usepackage{float}
\usepackage{breakurl}
\usepackage{booktabs}  

\newcommand{\todo}[1]{\textcolor{red}{\textbf{TODO:} #1}}
\newcommand{\info}[1]{\textcolor{blue}{\textbf{INFO:} #1}}
\newcommand{\del}[1]{\textcolor{gray}{\textbf{Asi smazat:} #1}}
\newcommand{\af}[1]{\textcolor{green}{\textbf{AF:} #1}}

\usepackage[T1]{fontenc}
\usepackage{graphicx}
\usepackage{color}

\begin{document}

\title{Hiding in Plain Sight: Query Obfuscation via Random Multilingual Searches}
%
%
\author{Anton Firc\inst{1}\and
Jan Klusáček\inst{1}  \and
Kamil Malinka\inst{1}}
\authorrunning{Firc et al.}

\institute{Brno University of Technology, Božetěchova 2, Brno, Czech Republic\\ \email{\{ifirc, malinka\}@fit.vut.cz}}

\maketitle              
\begin{abstract}
Modern search engines extensively personalize results by building detailed user profiles based on query history and behaviour. While personalization can enhance relevance, it introduces privacy risks and can lead to filter bubbles. This paper proposes and evaluates a lightweight, client-side query obfuscation strategy using randomly generated multilingual search queries to disrupt user profiling. Through controlled experiments on the Seznam.cz search engine, we assess the impact of interleaving real queries with obfuscating noise in various language configurations and ratios. Our findings show that while displayed search results remain largely stable, the search engine's identified user interests shift significantly under obfuscation. We further demonstrate that such random queries can prevent accurate profiling and overwrite established user profiles. This study provides practical evidence for query obfuscation as a viable privacy-preserving mechanism and introduces a tool that enables users to autonomously protect their search behaviour without modifying existing infrastructure.

\end{abstract}
\section{Introduction}

Online search engines are integral to how users interact with digital information. However, these systems often rely on extensive user profiling, leveraging demographic data, search histories, and behavioural signals to personalize search results and optimize advertisement delivery. While personalization can improve user experience, it raises well-documented concerns regarding privacy, surveillance, and information bias~\cite{bias,measurements}.

Personalized search can trap users in "filter bubbles," where only algorithmically curated content is visible. Worse, search engines often collect sensitive behavioural data without explicit user consent~\cite{ad_challanges,query_privacy}. These concerns have led to growing research on privacy-preserving search techniques, including anonymizing networks, private information retrieval (PIR), and query obfuscation~\cite{pir_protocols,query_privacy,privacy}.

Among these, query obfuscation presents a compelling trade-off: it requires only client-side changes, scales well, and can be deployed by privacy-conscious individuals. Tools like TrackMeNot have explored this concept by injecting decoy search traffic to obscure real interests~\cite{trackmenot}. However, prior work rarely evaluates the effectiveness of such methods under multilingual conditions or with respect to actual profile disruption over time.

This paper presents a practical approach to query obfuscation using multilingual random queries. We introduce a tool that interleaves real user queries with noise generated from language-specific dictionaries, simulating diverse user interests. Through controlled experiments using the Seznam.cz platform, which provides transparent user interest feedback, we evaluate the impact of obfuscation on search results and inferred user profiles.

The main contributions of this paper may be summarised as:
\begin{itemize}
    \item We introduce a lightweight tool for privacy-preserving query obfuscation using multilingual random queries designed for real-world usability.
    \item We empirically demonstrate that query obfuscation significantly alters user interest profiles, even when search result pages remain largely unaffected.
    \item We evaluate how language diversity and query ratios influence obfuscation effectiveness and show that profiles can be reshaped even after initial formation.
    \item We provide insights into the feasibility of user-controlled privacy enhancement without requiring changes to search engine infrastructure.
\end{itemize}

\sloppy
The work described in this paper results from a previously completed master's thesis~\cite{diplomka},
forming the core of this research. The implementation with additional materials is available at: \url{https://github.com/Sacek073/Protection-Against-Profiling-with-Random-Multilingual-Search}.


\section{Related Work}

Search personalization systems rely on detailed user profiling to deliver customized results. Profiling typically involves collecting and analyzing behavioural data such as search queries, click-through history, and location \cite{recommendation,trends,user_profile}. Profiles are built using explicit profiling (e.g., users providing information through forms such as age or interests), implicit profiling (e.g., inferring preferences from click behaviour), or a hybrid of both approaches \cite{bias}. While this personalization enhances user experience, it simultaneously creates severe privacy risks.

Search engines do not reveal their internal profiling algorithms, which hinders third-party analysis \cite{measurements}. Although patents such as \cite{patent} give some insight into legacy approaches, the current profiling mechanisms remain proprietary and continuously evolve.

\subsection{Privacy-Preserving Search Techniques}

To address privacy risks in web search, researchers have proposed various defence strategies categorized as \textit{Private Information Retrieval (PIR)}, \textit{anonymizing networks}, and \textit{query obfuscation} \cite{obfuscation,query_privacy}.

\textbf{Private Information Retrieval (PIR)} allows users to retrieve data without revealing their queries. Traditional PIR schemes are computationally expensive and assume server cooperation, which is incompatible with real-world search engines \cite{pir_protocols,privacy_agent}. Attempts to bypass these limitations include \textit{h(k)-PIR} schemes, which inject fake keywords into queries to achieve plausible deniability~\cite{hkpir}.

\textbf{Anonymizing networks} like Tor mask users' identities by routing traffic through multiple nodes. However, such networks still expose the query content to search engines and are susceptible to de-anonymization through query content analysis \cite{query_privacy,networks}. Studies demonstrate that attackers can re-identify users from anonymized pools using only short-term search histories and standard classifiers \cite{query_privacy}.

\textbf{Query obfuscation} is a client-side approach where fake queries are submitted alongside genuine ones to hide user interests \cite{trackmenot,obfuscation,query_privacy}. Tools like TrackMeNot exemplify this method by generating queries that mimic user interests based on prior search patterns. While this increases realism, it may inadvertently reinforce certain topics rather than obscure them. Moreover, TrackMeNot depends on browser extensions, lacks ongoing support, and provides limited control over the obfuscation strategy. In contrast, our approach operates independently using a headless browser, supports multilingual noise generation, and allows fine-tuning of query frequency and language diversity. These features enhance stealth, user autonomy, and the adaptability of the obfuscator in varied threat settings.

Recent work further reveals that many obfuscators—TrackMeNot included—are susceptible to filtering via semantic similarity and entropy-based models~\cite{web_privacy_formal_model}. Such findings highlight the need for more configurable and dynamic obfuscation tools, which our system addresses by incorporating randomized queries, multilingual context-switching, and evaluation grounded in profile inference feedback.

\subsection{Evaluation Frameworks and Limitations}

Recent studies propose frameworks to assess the effectiveness of obfuscation strategies systematically. OB-WSPES \cite{ob_evaluation} allows comparative analysis of obfuscation methods against modern attacks. Gervais et al. \cite{ccs_gervais} introduced a general methodology to evaluate privacy using query and semantic linkage metrics.

Newer approaches explore novel obfuscation mechanisms, such as differential privacy (DP), for query rewriting. For instance, Faggioli and Ferro demonstrate that DP-based obfuscation can achieve a tunable balance between privacy and relevance, outperforming traditional dummy-based methods under specific conditions \cite{dp_queries}.

Other proposals allow users to control the semantic distance and volume of fake queries, tailoring obfuscation strength to their privacy needs \cite{punagin2017}. These user-centric methods promote flexible trade-offs between utility and privacy but lack broad adoption or integration with real-world platforms.

\subsection{Positioning of This Work}

This study extends query obfuscation research by deploying a practical, multilingual obfuscator tested against real-time user profiling feedback from Seznam.cz. Unlike prior work, it directly evaluates how injected noise alters inferred user interests rather than only focusing on query-level traceability. This contributes empirical insights to an underexplored dimension of user-centric privacy tools. By enabling fine-grained control over language selection, timing, and query volume, our tool also provides a more adaptable and modular platform than existing solutions like TrackMeNot.

\section{Threat Model and Assumptions}

We consider a \textbf{passive profiling adversary}, modelled as the search engine itself. It collects and processes user queries to infer interests for personalization and advertising. The adversary has full access to submitted queries and session metadata but does not control the user's device or actively interfere with the content.

We aim to degrade the accuracy of inferred user profiles by injecting randomized, multilingual decoy queries. The adversary is not assumed to detect or classify obfuscated queries in real-time, though it may apply generic profiling algorithms over aggregated data.

The evaluation is conducted on Seznam.cz, which openly displays user interest profiles. This enables empirical observation of how obfuscation impacts profiling outcomes. We do not defend against semantic or behavioural de-anonymization attacks, focusing instead on practical, client-side resistance to standard profiling mechanisms.

\paragraph*{Adversarial Capabilities.}
Although we model the search engine primarily as a passive profiler, real-world systems may deploy active or semi-active mechanisms to detect obfuscation attempts. These include~\cite{web_privacy_formal_model,query_privacy}:
\begin{itemize}
    \item \textbf{Temporal anomalies} — Detection of unnatural regularity in query timing or volume.
    \item \textbf{Semantic incoherence} — Recognition of disjointed or unrelated topics inconsistent with typical user interests.
    \item \textbf{Language switching patterns} — Multilingual behaviour without corresponding context changes (e.g., UI language).
    \item \textbf{Lack of engagement} — Missing interaction signals such as clicks or dwell time after queries.
\end{itemize}

Our tool addresses some of these issues via randomized inter-query delays, plausible query lengths, and headless browser automation that mimics human interaction patterns. However, a powerful adversary could still filter out low-quality or suspicious queries post hoc, especially in commercial settings optimized for personalization.

We restrict our evaluation to scenarios where such detection is not yet deployed or is not sensitive enough to filter our approach reliably. Future work may extend this model to account for stronger adversaries with access to richer behavioural telemetry or multi-session fingerprinting techniques.

\section{Obfuscation Tool Architecture and Workflow}

We developed a custom automation tool to evaluate the effectiveness of multilingual query obfuscation, as existing solutions such as TrackMeNot~\footnote{\url{https://www.trackmenot.io/}} lacked required features and long-term maintainability~\cite{trackmenot,query_privacy}. The tool simulates user-like search behaviour by issuing genuine and randomized multilingual queries to Seznam.cz. It enables controlled experimentation on profiling disruption in a real-world environment where inferred interests are visible to the user.

The tool is implemented using Puppeteer, a Node.js library for browser automation, and extended with the \texttt{puppeteer-extra-plugin-stealth}\footnote{\url{https://www.npmjs.com/package/puppeteer-extra-plugin-stealth}} to mimic natural user interaction. It operates headless, managing login, query issuance, and result collection while minimizing detectability. All actions, including delays and query typing, emulate genuine usage patterns.

\subsection{Search Generation}

Obfuscating queries are generated from precompiled language-specific dictionaries comprising general vocabulary chosen to represent broad, demographically diverse topics. To ensure randomness and topic dilution, query length and language rotation follow user-defined configurations. Although ideally, the tool would simulate diverse demographic profiles, this was deemed infeasible; hence, randomness from broad lexical sources was prioritized.

\subsection{Modes of Operation}

The tool operates in two distinct modes:
\begin{itemize}
    \item \texttt{the\_tool}: Issues randomized queries using selected language dictionaries. Users configure the query length range and language pool. Queries are interleaved in a round-robin fashion to simulate multilingual behaviour.
    \item \texttt{queries}: Executes predefined queries from a provided list intended for targeted profiling or evaluation scenarios. This mode captures search results to measure post-query personalization.
\end{itemize}

\subsection{Workflow Summary}

The tool's workflow includes:
\begin{enumerate}
    \item Launching a stealth-configured headless browser.
    \item Logging into a user account on Seznam.cz.
    \item Loading either language wordlists or predefined queries.
    \item Executing queries in a loop: entering text, triggering search, collecting result metadata, and sleeping between iterations.
    \item Saving structured outputs in JSON format containing query, language, user, timestamp, and result objects with links and categorization.
\end{enumerate}

To handle unstable elements such as malformed result pages or missing CSS selectors, all browser operations are wrapped in recovery blocks that log errors and resume operation. This ensures robustness against transient failures during long-term operation.

\subsection{Output Format}

Search outputs are stored in JSON format for downstream evaluation. Each entry includes the query issued, the language used, the user ID, the timestamp, and a list of result objects. Results are further annotated with a boolean flag indicating whether the result is a genuine external link or a platform-specific element (e.g., videos, image results, local services), enabling finer-grained analysis of content types returned during obfuscated and targeted searches.

\section{Experimental Setup}
\label{sec:experiments}

The goal of this study is to evaluate the effectiveness of multilingual query obfuscation in disrupting search engines' user profiling. We investigate this through four controlled experiments designed to answer the following research questions:

\noindent \textbf{RQ1:} Does multilingual query obfuscation alter displayed results or inferred user interests?\\
\noindent \textbf{RQ2:} How does the number of languages used in random queries affect profiling?\\
\noindent \textbf{RQ3:} How does varying the ratio of genuine to random queries influence outcomes?\\
\noindent \textbf{RQ4:} Can obfuscation reshape an already formed user profile?

\subsection{System Configuration}

All experiments were conducted using Virtual Machines (VMs) hosted on a single physical machine within a university subnet to minimize geolocation bias and temporal noise due to network routing or infrastructure differences~\cite{measurements}. The VMs were isolated using distinct user accounts and browser instances, ensuring independent sessions with no cookie sharing, cross-account leakage, or fingerprint carryover. Browser cache and history were cleared at the beginning of each experiment.

Three base VMs were created:
\begin{itemize}
    \item \textbf{Normal VM}: submits only profiling queries.
    \item \textbf{Control VM}: identical to Normal, used for baseline comparison.
    \item \textbf{Tool VM}: submits profiling queries and runs background random multilingual queries.
\end{itemize}

Each VM had a manually registered Seznam.cz account with identical demographic information (male, born 01/01/2000). No other personalization was applied. A dedicated prepaid SIM card was used to activate all accounts, avoiding any linkability to prior identity or external services.

\subsection{Query Design and Execution}

To simulate realistic user behaviour, profiling queries were constructed to reflect an interest in three categories: \textit{sports}, \textit{technology}, and \textit{travel}. A total of 300 Czech-language queries were crafted (100 per category), spread evenly across 10 daily batches. Sample queries include: \textit{"zimní olympiáda", "stackoverflow", "letenky online"}. Each VM submitted 30 queries per day (one every 960 seconds), mimicking natural inter-search intervals observed in human users.

The Tool VM additionally submitted 90 background queries daily, generated from dictionaries in Czech, English, French, and Spanish. These dictionaries were curated to ensure topic diversity and realism. Queries were constructed by selecting 1–3 random words inspired by research on natural query lengths in real-world search behavior~\cite{brevity}. Random searches were spaced at 320-second intervals, resulting in a 3:1 ratio of random to genuine queries. These delay values were chosen to maintain operational stealth and to minimize the carry-over effect~\cite{measurements}, where rapid consecutive queries could influence search results.

All experiments were run for 8-hour periods across 10 days. Queries were issued sequentially via a headless browser controlled using Puppeteer, configured with randomized delays, human-like typing simulation, and stealth plugins to avoid detection by bot-detection heuristics.

\subsection{Experiments}

\noindent \textbf{Experiment 1: Baseline Effectiveness.} The Tool VM submitted profiling and obfuscation queries in parallel to test whether the additional traffic disrupted interest profiling or affected search results.

\noindent \textbf{Experiment 2: Language Diversity.} This experiment introduced:
\begin{itemize}
    \item \textbf{Language Low VM}: uses only Czech for random queries.
    \item \textbf{Language High VM}: uses eight languages, including Czech, English, French, Italian, Slovak, Spanish, Turkish, and Ukrainian.
\end{itemize}
The goal was to assess whether broader linguistic diversity improves obfuscation. Both used a 1:3 profiling-to-random ratio.

\noindent \textbf{Experiment 3: Query Ratio.} This experiment varied the volume of random queries:
\begin{itemize}
    \item \textbf{Ratio Low VM}: 1:1 ratio (equal number of genuine and random queries).
    \item \textbf{Ratio High VM}: 1:7 ratio (one genuine query to seven obfuscated ones).
\end{itemize}
This aimed to evaluate how dilution strength affects personalization.

\noindent \textbf{Experiment 4: Delayed Obfuscation.} The \textbf{Delay VM} performed only profiling queries for the first 5 days, then submitted random queries to evaluate if pre-established profiles could be altered retrospectively.

\subsection{Search Engine Selection}

Seznam.cz was selected due to its transparent profiling model and lack of restrictions on automated access. Unlike Google or Bing, Seznam.cz displays daily-updated “Areas of Interest” derived from user search history, which users can view or delete at \url{https://ucet.seznam.cz/activity/targeting}. This feature allows direct monitoring of profile evolution over time, making it uniquely suited for controlled obfuscation studies.

\subsection{Evaluation Metrics and Analysis}

To assess changes in personalization and content, two metrics were used:
\begin{itemize}
    \item \textbf{Jaccard Index}~\cite{jaccard}: measures set similarity between result or interest sets; ranges from 0 (no overlap) to 1 (identical).
    \item \textbf{Edit Distance}~\cite{levenshtein}: quantifies reordering and substitution cost between ranked lists.
\end{itemize}

All profiling queries captured the top-10 search results. Interest profiles were extracted daily from the Seznam interface. Comparisons were made between VM pairs (e.g., Normal vs. Tool) across days. Statistical significance was tested using the Shapiro-Wilk test for normality~\cite{normality}. As no datasets followed a normal distribution, non-parametric Mann-Whitney U tests~\cite{nonparametric} were used with $\alpha = 0.05$ to determine significant differences.

\section{Results}
\label{sec:results}

The experiments were conducted sequentially over more than 40 days. Search results and identified interests were collected daily and evaluated using the Jaccard Index and Edit Distance. Statistical significance was assessed using the Mann-Whitney U test ($\alpha = 0.05$). Table~\ref{tab:p-values} summarizes the p-values for all experiments.

\begin{table}[!htbp]
    \centering
    \caption{P-values from the Mann-Whitney U test for search results and identified interests.}
    \begin{tabular}{lcc|cc}
        \toprule
        \textbf{Experiment} & \multicolumn{2}{c|}{\textbf{Search Results}} & \multicolumn{2}{c}{\textbf{Identified Interests}} \\
        & \emph{Jaccard} & \emph{Edit Dist.} & \emph{Jaccard} & \emph{Edit Dist.} \\
        \midrule
        One             & 0.0018 & 0.0222 & 0.0001 & 0.0001 \\
        Two (Low)       & 0.0803 & 0.1544 & $<$ 0.0001 & $<$ 0.0001 \\
        Two (High)      & 0.0733 & 0.1324 & $<$ 0.0001 & $<$ 0.0001 \\
        Three (Low)     & 0.5173 & 0.4955 & $<$ 0.0001 & $<$ 0.0001 \\
        Three (High)    & 0.1108 & 0.1617 & $<$ 0.0001 & $<$ 0.0001 \\
        Four            & 0.4061 & 0.1568 & 0.0402 & 0.0402 \\
        \bottomrule
    \end{tabular}
    \label{tab:p-values}
\end{table}

\subsection*{RQ1: Effectiveness of Multilingual Obfuscation}

\begin{figure}[!htbp]
    \centering
    \includegraphics[width=0.8\textwidth]{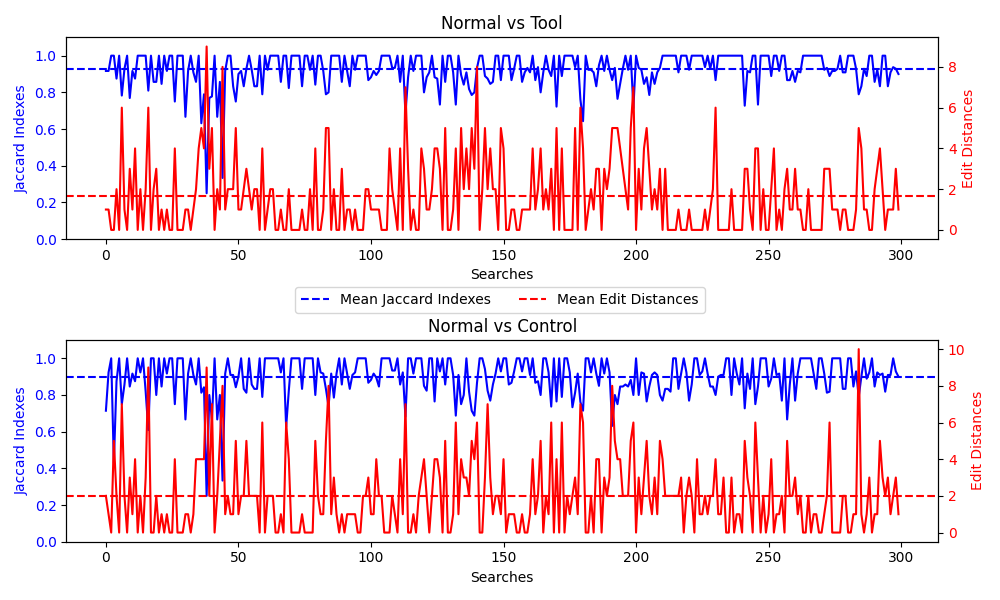}
    \caption{Search Result Similarity (Jaccard Index and Edit Distance).}
    \label{normal_searches}
\end{figure}

\begin{figure}[!htbp]
    \centering
    \includegraphics[width=0.8\textwidth]{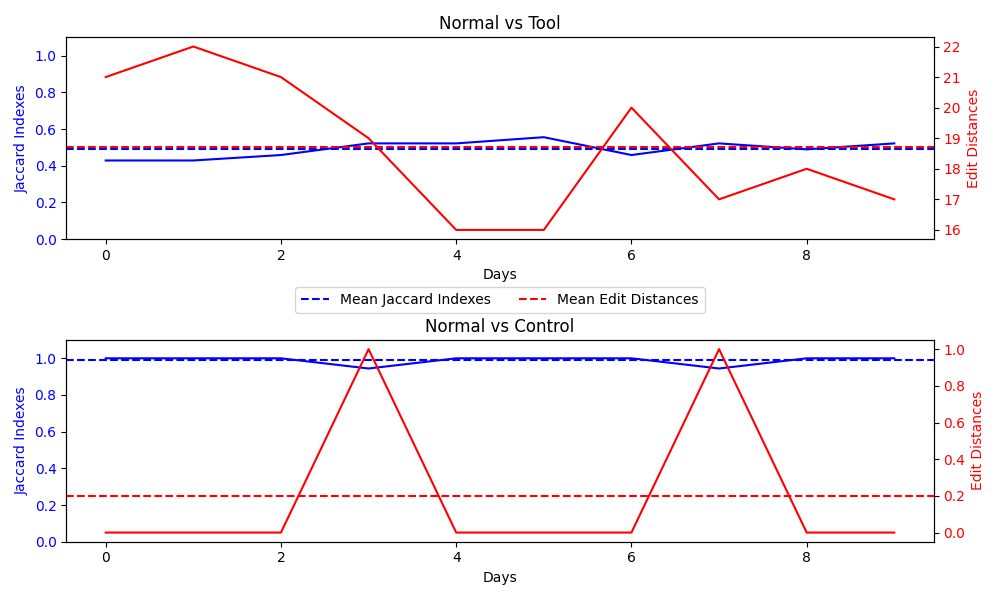}
    \caption{Identified Interests Similarity. Note the Edit Distance scale.}
    \label{normal_interests}
\end{figure}

Experiment One showed statistically significant differences between the Tool and Normal VMs in both search results and identified interests. Although changes in search results were modest (Figure~\ref{normal_searches}), the differences in identified interests were pronounced (Figure~\ref{normal_interests}). The Tool VM’s interests had consistently lower Jaccard values and higher Edit Distances, suggesting successful disruption of profiling.

\subsection*{RQ2: Language Diversity Impact}

\begin{figure}[!htbp]
    \centering
    \includegraphics[width=0.8\textwidth]{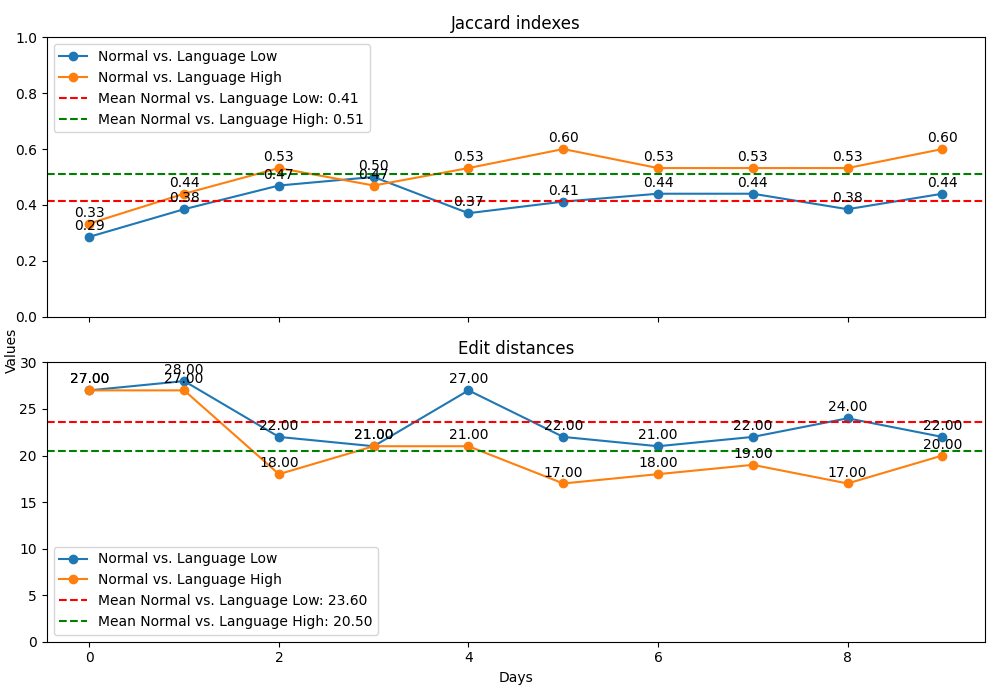}
    \caption{Language Experiment: Identified Interests (Jaccard Index - Top, Edit Distance - Bottom).}
    \label{language}
\end{figure}

In Experiment Two, the number of languages used in random queries had minimal effect on search results but a significant influence on identified interests (Figure~\ref{language}). Using only Czech yielded more divergence from the Normal VM than using eight languages, suggesting that less linguistic diversity may better obscure interest profiles.

\subsection*{RQ3: Obfuscation Ratio Effects}

\begin{figure}
    \centering
    \includegraphics[width=0.8\textwidth]{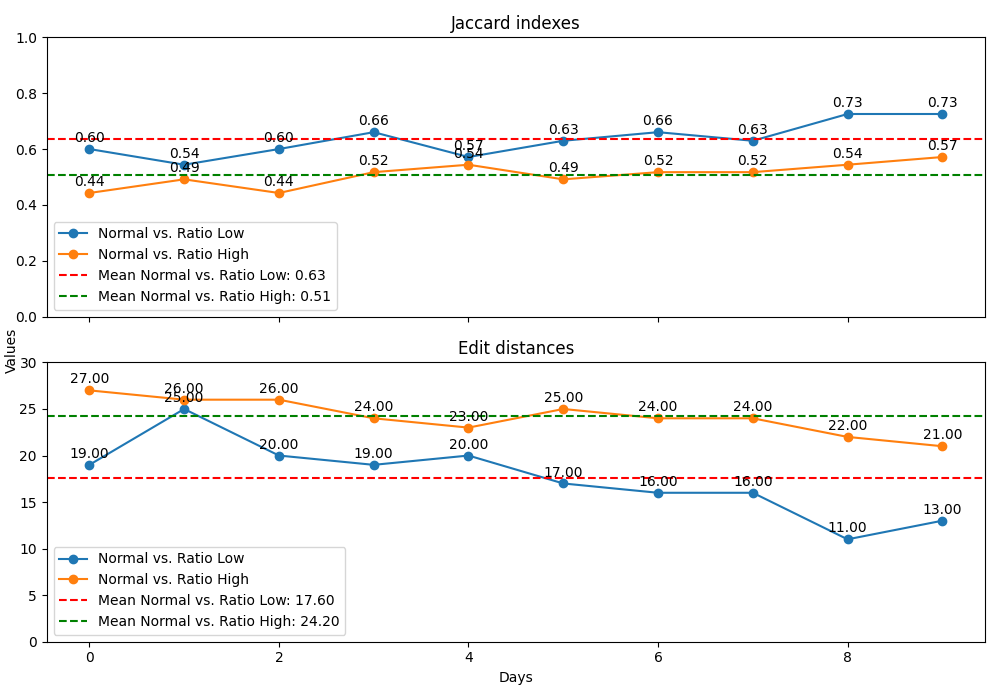}
    \caption{Ratio Experiment: Identified Interests (Jaccard Index - Top, Edit Distance - Bottom).}
    \label{ratio}
\end{figure}

Experiment Three revealed no statistically significant changes in search results, but strong effects on identified interests. A higher random-to-genuine query ratio (1:7) led to greater divergence from the Normal VM compared to a lower ratio (1:1), as illustrated in Figure~\ref{ratio}. This confirms that stronger obfuscation correlates with better profile disruption.

\subsection*{RQ4: Overwriting Existing Profiles}

\begin{figure}
    \centering
    \includegraphics[width=0.9\textwidth]{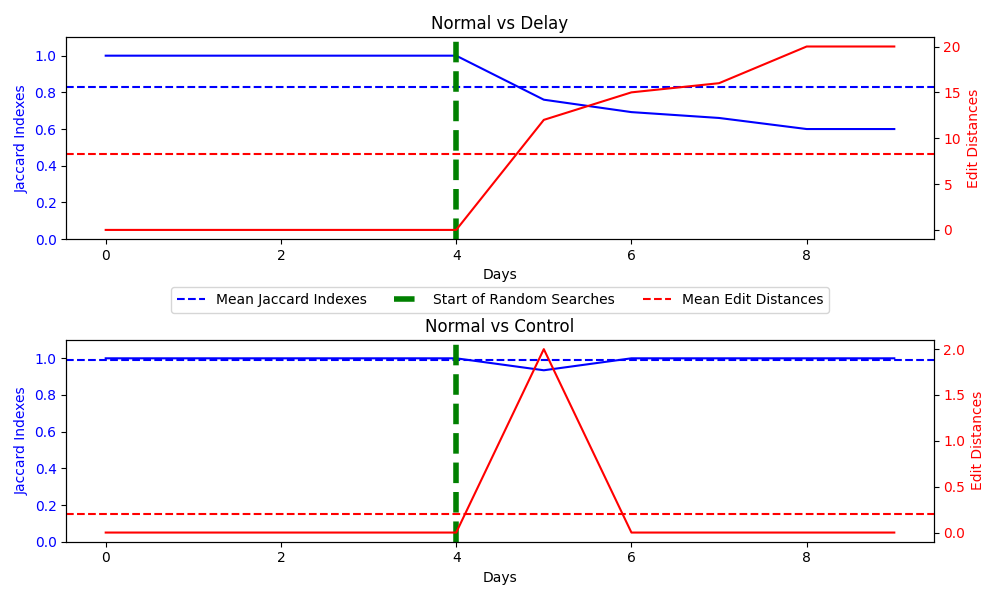}
    \caption{Delayed Obfuscation: Changes in Identified Interests.}
    \label{delay_interests}
\end{figure}

In Experiment Four, the Delay VM began random queries after five days of profiling. Although search results remained largely unchanged, a shift in identified interests occurred after random queries started (Figure~\ref{delay_interests}). Edit Distance increased and Jaccard Index decreased, indicating successful profile mutation.

\section{Discussion}

This study provides empirical evidence that query obfuscation—particularly through multilingual random queries—can disrupt user profiling mechanisms used by search engines. Across multiple experiments, we observed consistent alterations in inferred user interests, even when search results remained largely stable. These findings validate the premise that obfuscation-based methods can be practical tools for enhancing search privacy, especially in user-centred, client-side deployments.

\subsection{Impact on Profiling vs. Search Results}

A key insight from our experiments is the decoupling between personalized search results and user profiling. While obfuscation had a limited impact on the top search results returned for specific queries, it significantly affected the interest profiles assigned to users. This suggests that search personalization pipelines operate on a longer-term behavioural model, where interest inference is more sensitive to volume and diversity of query history than short-term result ranking.

The implications are twofold. First, users concerned with behavioural profiling—e.g., for advertising or content filtering—can benefit from lightweight obfuscation without compromising search result usability. Second, this dissociation between visible results and backend profiling highlights the opacity of personalization systems and the need for tools that let users intervene in shaping or obscuring their inferred identities.

\subsection{Strategic Obfuscation: Language and Volume}

The experiments examining language diversity and obfuscation ratios reveal a nuanced view of obfuscation strategy. Surprisingly, obfuscation using only Czech yielded greater disruption than multilingual approaches, likely due to the profiling engine's focus on Czech-language content. This suggests that effective obfuscation must account for the language model and cultural domain of the search engine in use. Simply increasing linguistic variety may introduce noise without undermining the assumptions of the underlying profiling algorithms.

Additionally, increasing the volume of random queries consistently improved concealment. A 1:7 ratio of genuine to random queries yielded the most privacy-preserving profiles. This supports the conceptual intuition that in profiling systems where relevance is inferred from frequency and recency, flooding the model with decoy data can obscure meaningful signals. This finding aligns with adversarial obfuscation literature and confirms that even simple random query strategies can introduce sufficient entropy to distort user modelling.

\subsection{Dynamic Reprofiling: Reversibility of Interests}

A particularly promising outcome was the ability to reshape an already established interest profile. In Experiment Four, delayed obfuscation reversed the trajectory of interest inference, leading to measurable divergence from the pre-existing profile. This finding has practical significance: users who are already heavily profiled are not locked into their inferred identities. Even without deleting accounts or histories, profile trajectories can be perturbed through non-invasive means.

This also raises an important ethical and design question—should users have the ability to audit and intervene in their profiles more actively? Search engines currently do not offer this capability, but our findings suggest it is both feasible and technically low-cost.

\subsection{Positioning Within the Privacy Landscape}

This work contributes to the growing class of user-controllable, zero-trust privacy mechanisms. Unlike Private Information Retrieval (PIR), which requires cooperation from service providers, and anonymizing networks like Tor, which protect identity but not content, our tool defends against profiling by polluting the data corpus used for inference. It is a pragmatic middle-ground solution: usable, scalable, and compatible with current web infrastructure.

Moreover, the use of Seznam.cz provides a unique empirical window into profiling processes that are often opaque on platforms like Google. While this limits generalizability, it provides rare observable feedback on how query history shapes user models in practice.

\subsection{Risks and Evasion Potential}

That said, these techniques are not foolproof. Our tool assumes a non-adversarial search engine that does not actively filter or flag suspicious query patterns. In real-world deployments, engines may develop classifiers to identify and suppress obfuscation traffic. Semantic and behavioural fingerprinting (e.g., dwell time, query structure, device identifiers) could also be used to infer query authenticity.

Additionally, overuse of obfuscation may degrade personalization quality for users who value it. While some users may accept this trade-off for privacy, others may seek selective control over which queries are masked. This presents opportunities for future work in adaptive obfuscation—balancing utility and privacy dynamically based on context or sensitivity.

\subsection{Limitations}

Several limitations should be acknowledged. The use of Seznam.cz, while justifiable from a legal and operational standpoint, constrains broader applicability. Its user base, infrastructure, and profiling logic differ from global engines. The Czech language and localized query structures may have biased the obfuscation results.

Furthermore, automation limits were discovered empirically due to a lack of documentation. These constraints affected query throughput and forced serialized experiment execution, possibly introducing day-to-day search index drift.

Finally, while VMs were synchronized within minutes, they could not issue queries simultaneously. This introduces potential temporal noise, although prior work suggests such variation is minimal when comparing interest profiles rather than instantaneous result rankings.

\subsection{Toward Controllable Search Privacy}

Despite these limitations, this work shows that personalization and profiling can be manipulated without breaking terms of service, compromising search engines, or deploying heavyweight cryptographic tools. The results suggest a future where users actively manage their digital shadows—not by opting out entirely, but by introducing ambiguity into the systems that seek to define them.

\subsection{Usability and Deployment Considerations}

From a deployment perspective, our tool is lightweight and can run passively in the background during user browsing sessions. However, large-scale use may face challenges such as detection by advanced bot-filtering systems or throttling based on abnormal traffic patterns. Furthermore, while the tool simulates natural search behavior, it does not currently simulate user engagement (e.g., clicks or scrolling), which may limit realism in adversarial settings. Integrating feedback-based adaptation or browser-based query diversification may improve stealth without significantly increasing user-side complexity.

\subsection{Use Case: Lightweight Daily Obfuscation}

Consider a privacy-aware user performing 10 genuine searches daily. Using our tool in the background during browsing hours (e.g., 9:00–17:00), the system automatically injects random multilingual queries at a 1:3 ratio. This lightweight activity runs in a headless browser, mimicking typical user behaviour without disrupting the user’s experience.

Over time, the search engine’s profile is diluted with noise, reducing personalization accuracy. This approach requires no server-side changes and minimal resources, and it aligns with terms of service, making it practical for everyday privacy protection.

\section{Conclusion}

This study demonstrates that random query obfuscation can effectively disrupt user profiling in search engines, particularly by altering inferred interests without noticeably impacting search results. Through experiments conducted on Seznam.cz, we showed that using a single language (Czech) and a high ratio of random to genuine queries offers strong obfuscation performance. Furthermore, obfuscation remains effective even when applied to previously established profiles, enabling retrospective privacy protection.

Our approach is lightweight, user-friendly, and compliant with search engine terms of service. It can be deployed in the background during regular browsing sessions, offering practical privacy enhancements with minimal user intervention.

Future research should investigate optimal obfuscation strategies, explore applicability to broader personalization contexts, and assess long-term effects and integration into real-world privacy tools.

Although the experiments were conducted on Seznam.cz due to its transparency and terms of service, the fundamental mechanism of query-based profiling is shared by most major search engines. Consequently, the observed effects of obfuscation are expected to transfer to platforms like Google or Bing, albeit with variations due to their more complex personalization models. Testing across these platforms remains an important direction for future research, ideally leveraging ethical frameworks or approved APIs.

\subsubsection{\ackname}
This work was supported by the Brno University of Technology internal project FIT-S-23-8151.


%
%
%
\bibliographystyle{splncs04}
\bibliography{bibliography}

\end{document}